\documentclass[aps,prb,amsmath,amssymb,
reprint,%
superscriptaddress,
]{revtex4-1}

\usepackage{graphicx}% Include figure files
\usepackage{subfigure}
\usepackage{textgreek}
\usepackage{ulem}
\usepackage{color}

\usepackage[latin1]{inputenc}
\usepackage[T1]{fontenc}

\begin{document}

\title[Ultrafast Band Structure Dynamics in Bulk 1{\it T}-VSe$_2$]
{Ultrafast Band Structure Dynamics in Bulk 1{\it T}-VSe$_2$}

\author{Wibke Bronsch}
\email[Corresponding author: ]{wibke.bronsch@elettra.eu}
\affiliation{Elettra - Sincrotrone Trieste S.C.p.A., Strada Statale 14 - km 163.5 in AREA Science Park, 34149 Basovizza, Trieste, Italy}

\author{Manuel Tuniz}
\affiliation{Dipartimento di Fisica, Universit\`{a} degli Studi di Trieste, 34127 Trieste, Italy}

\author{Denny Puntel}
\affiliation{Dipartimento di Fisica, Universit\`{a} degli Studi di Trieste, 34127 Trieste, Italy}

\author{Alessandro Giammarino}
\affiliation{Dipartimento di Fisica, Universit\`{a} degli Studi di Trieste, 34127 Trieste, Italy}

\author{Fulvio Parmigiani}
\affiliation{Dipartimento di Fisica, Universit\`{a} degli Studi di Trieste, 34127 Trieste, Italy}
\affiliation{Elettra - Sincrotrone Trieste S.C.p.A., Strada Statale 14 - km 163.5 in AREA Science Park, 34149 Basovizza, Trieste, Italy}
\affiliation{International Faculty, University of Cologne, Albertus-Magnus-Platz, 50923 Cologne, Germany}

\author{Yang-hao Chan}
\affiliation{Institute of Atomic and Molecular Sciences, Academia Sinica, Taipei 10617, Taiwan}
\affiliation{Physics Division, National Center of Theoretical Physics, Taipei 10617, Taiwan}

\author{Federico Cilento}
\email[Corresponding author: ]{federico.cilento@elettra.eu}
\affiliation{Elettra - Sincrotrone Trieste S.C.p.A., Strada Statale 14 - km 163.5 in AREA Science Park, 34149 Basovizza, Trieste, Italy}

\date{\today}

\begin{abstract}
Complex materials encompassing different phases of matter can display new photoinduced metastable states differing from those attainable under equilibrium conditions. 
These states can be realized when energy is injected in the material following a non-equilibrium pathway, unbalancing the unperturbed energy landscape of the material. 
Guided by the fact that photoemission experiments allow for detailed insights in the electronic band structure of ordered systems, here we study bulk 1{\it T}-VSe$_2$ in its metallic and charge-density-wave phase by time- and angle-resolved photoelectron spectroscopy.
After near-infrared optical excitation, the system shows a net increase of the density of states in the energy range of the valence bands, in the vicinity of the Fermi level, lasting for several picoseconds. 
We discuss possible origins as band shifts or correlation effects on the basis of a band structure analysis.
Our results uncover the possibility of altering the electronic band structure of bulk 1{\it T}-VSe$_2$ for low excitation fluences, contributing to the understanding of light-induced electronic states.
\end{abstract}

\maketitle

\section{introduction}
\begin{figure}[h!]
\includegraphics[width=8.75cm]{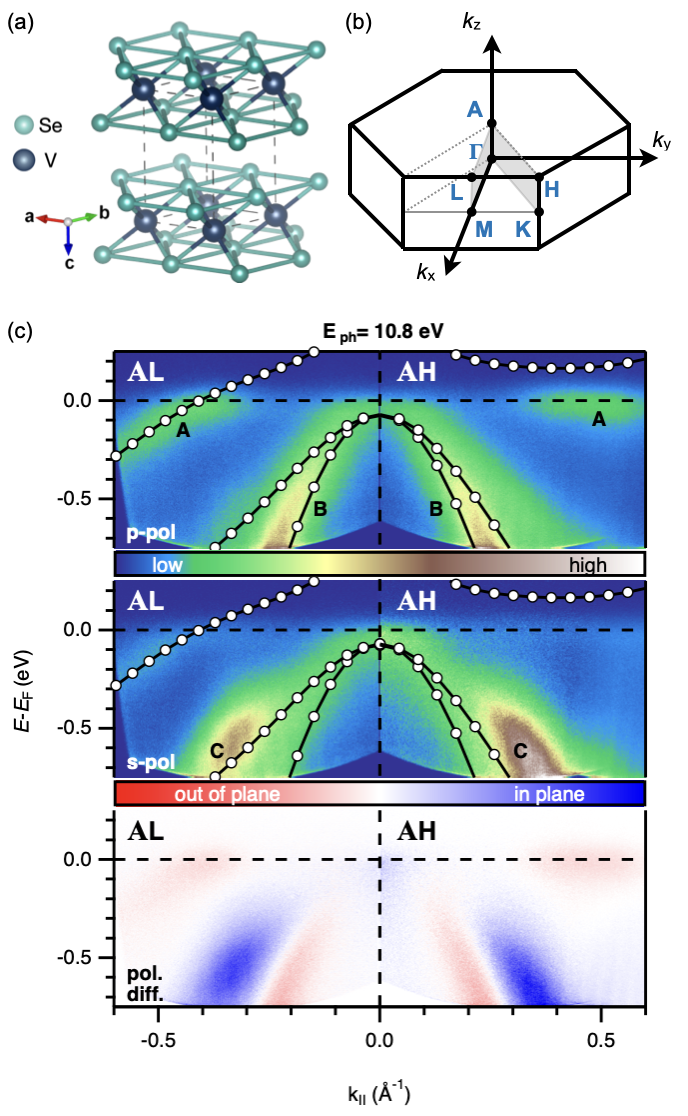}
\caption{
Bulk crystal structure (a) and hexagonal Brillouin zone (b) of 1{\it T}-VSe$_2$. (c) ARPES data recorded with 10.8\,eV probe light in $p$- ({\bf top}) and $s$-polarization ({\bf middle}) along the AL (left) and the AH (right) direction of the BZ. Three different bands, labelled as A, B and C, are revealed. The {\bf bottom} panel shows a difference plot of the top and middle panel ($p$-pol - $s$-pol). Red color indicates states with higher intensity when probed with $p$-polarized light and blue predominantly probed by $s$-polarized light. 
We compare our data with DFT band structure calculations in the constant-$k_z$ plane intersecting the A point of the BZ. The calculations were shifted by 50\,meV towards the Fermi level to match the position of band A in the AL direction.
}
\label{11eV_ARPES}
\end{figure}

Layered transition-metal dichalcogenides (TMDCs) are attracting particular attention since they exhibit a variety of physical properties such as spin- and valley-coupled two-dimensional superconductivity, exciton Hall effect and exotic charge-density-wave (CDW) transitions.\cite{Manzeli2017}
Path breaking works on light induced CDWs in compunds as VO$_2$\cite{Morrison2014,Muraoka2018} and LaTe$_3$ \cite{Kogar2020} triggered the search for similar effects in TMDCs.
Photoexcitation can drive TMDC-CDW systems in metastable states surviving for several picoseconds, as reported for 1$T$-TiSe$_2$\cite{Hedayat2019,Duan2023}, 1$T$-TaSe$_2$\cite{Shi2019} and 1$T$-VSe$_2$\cite{Majchrzak2021}. 
Origins of these states are attributed to phonon bottleneck effects and optically-induced phase transitions from the CDW phase into a metallic phase.
Duan {\it et al.} discovered that for TiSe$_2$ an optically-induced metastable metallic phase can be realized far below the CDW transition temperature. 
This state, showing different properties than the equilibrium normal phase \cite{Duan2023}, was connected to the optical triggering of a halted motion of the atoms via coherent electron-phonon coupling processes.
In the case of 1{\it T}-VSe$_2$, instead, phonon bottleneck effects after optical excitation in its metallic phase were discussed.\cite{Majchrzak2021} 
Recent interest has been drawn to the monolayer limit of VSe$_2$ due to the enhanced CDW transition temperature and enlarged band gap.\cite{Biswas2021,Umemoto2019,Feng2018,Duvjir2018} 
Biswas {\it et al.}\cite{Biswas2021} report on ultrafast light induced insulator-metal transition in 2D VSe$_2$.
However, there are still open questions regarding the bulk properties of VSe$_2$ to be answered, to gain a profound understanding of the physical mechanisms that determine its properties and phase transitions.
With this work we contribute to the understanding of the optically-induced metastable state of 1{\it T}-VSe$_2$ by performing time- and angle-resolved photoemission (tr-ARPES) experiments. 
We excite the material with 1.55\,eV pump pulses and probe it at its $\Gamma$ point with $\approx$\,200\,fs time and $\approx$\,50\,meV energy resolution.
Comparing results acquired at three different sample temperatures (well above, around and well below the CDW transition temperature of T$_{\rm c}=$110\,K\cite{Diego2021}), we also address the influence of the initial phase of the system on the observed metastable modification.
\\
ARPES gives access to the valence band structure of a material, which in the case of 1{\it T}-VSe$_2$ is given by the vanadium 3d (V$_{\rm 3d}$) and selenium 4p (Se$_{\rm 4p}$) bands.\cite{Terashima2003,Strocov2012,Majchrzak2021}
Beyond the equilibrium band structure measurements, tr-ARPES allows studying the material response to an optical stimulus that drives the system far from equilibrium. 
This technique allows studying band-specific carrier dynamics which reveal {\it e.g.} information about the electron-electron and electron-phonon interactions in the system under investigation. %\cite{XXX} 
Moreover, tr-ARPES was shown to be an adequate technique to study light-induced metastable states in the class of TMDC materials and beyond.\cite{Majchrzak2021,Fanfarillo2021,Duan2023,Hedayat2019,Shi2019}
Here, we are presenting band structure measurements on bulk 1{\it T}-VSe$_2$ for different probe polarizations, which allows us to experimentally disentangle the contributions of the vanadium and selenium bands in the valence band structure, exploiting the selective access to bands with different orbital symmetry as offered by a different probe polarization.
% to investigate the electron-phonon coupling ...
In Fig.s\,\ref{11eV_ARPES}(a),(b) we show the layered crystal structure of bulk 1{\it T}-VSe$_2$ and its hexagonal Brillouin zone (BZ). 
The two photon energies used in this work, namely 6.2\,eV and 10.8\,eV, probe the three dimensional band structure in the AHL and $\Gamma$KM plane, respectively.

\section{Methods}
\subsection{Photoelectron spectroscopy}
ARPES measurements at equilibrium were performed using a photon energy of 10.8\,eV. 
The angle of incidence of the laser beam is 30$^{\circ}$ with respect to the surface normal. 
The polarization of the 10.8\,eV photons is controlled by varying the polarization of the incoming beam in the last harmonic upconversion step, which is a third-harmonic generation stage performed in a cell filled with Xe gas. More details about the setup are reported in Ref.\,[\onlinecite{Peli2020}].
Due to the measurement geometry $p$-polarized light always keeps a component parallel to the sample surface.
\\
Tr-ARPES experiments were performed using the 800\,nm (1.55\,eV) fundamental beam of an amplified short pulse Ti:sapphire laser system as a pump beam and its 4th harmonic as a probe. 
The polarization of the probe beam is varied by a $\lambda/2$ wave plate.
\\
All data were acquired with a SPECS PHOIBOS 225 hemispherical electron analyzer, equipped with a delay line detector by Surface Concept. We recorded all data in the wide-angle mode of the spectrometer.
\\\\
Clean sample surfaces were prepared by cleaving {\it in situ} a bulk VSe$_2$ crystal, purchased from HQ Graphene, using a kapton tape. 
The background pressure of the ultra-high vacuum chamber, in which we were cleaving the samples was $\sim 5\cdot10^{-10}$\,mbar, while the measurements were performed at a background pressure of $<3\cdot10^{-10}$\,mbar.
The required orientation of the samples was set via information from \textit{in-situ} low-energy electron diffraction measurements.

\subsection{Density functional theory calculations}
Density functional theory calculations were performed using the Vienna ab initio package (VASP)\cite{Kresse1993,Kresse1996a,Kresse1996b} with the projector augmented wave method\cite{Bloechl1994,Kresse1999}. 
Freestanding films were modeled with a 16\,\AA\,vacuum gap for the supercell. 
A plane-wave energy cutoff of 320\,eV and a 16x16x10 $k$-mesh were adopted for the films.
The generalized gradient approximation (GGA) with the Perdew-Burke-Ernzerhof (PBE) functional\cite{Perdew1996} was used; the bulk lattice constants are $a = 3.35$\,\AA\,and $c = 6.1$\,\AA. 
Spin-orbit interactions were incorporated throughout the computations.
For joint-density of states calculations (cf. supplementary material) we use Wannier90 package.\cite{Pizzi2020}

\section{results}
\subsection{Polarization dependent ARPES at equilibrium}
The valence band structure of 1{\it T}-VSe$_2$ originates from the V$_{\rm 3d}$ and the Se$_{\rm 4p}$ bands. 
According to band structure calculations, these bands show different dispersion along the $\Gamma$A direction of its hexagonal BZ.\cite{Jolie2019,Majchrzak2021,Falke2021,Henke2020} 
Whereas four of those bands overlap at the $\Gamma$ point of the BZ, at the A point they are separated from each other.
ARPES at different photon energies and light polarizations, makes these bands distinguishable by their different dispersion in the three dimensional momentum space and orbital band character.
DFT band structure calculations reported clear changes in the orbital character of the bands, despite strong hybridizations in certain regions of the BZ.\cite{Majchrzak2021,Jolie2019}
In the following we will compare ARPES measurements for different light polarizations to demonstrate that different parts of the valence band structure can be emphasized selectively with $s$- and $p$-polarized light. 
\\
Figure\,\ref{11eV_ARPES}(c) shows the valence band structure of bulk VSe$_2$, as probed at room temperature at 10.8\,eV photon energy.
Considering the $k_{\rm z}$ dispersion of the valence bands and the inner potential of 16 eV reported by Falke {\it et al.} \cite{Falke2021}, we conclude that in the hypothesis of free-electron final states we are probing the sample band structure close to the A-point of the BZ. 
We map the photoemission intensity as a function of binding energy with respect to the Fermi level ($E-E_{\rm F}$) and electron momentum along the AL and AH direction ($k_{\parallel}$), respectively.
The uppermost panel shows the result acquired with $p$-polarized light.
We distinguish two bands, one positioned at the Fermi level with a flat dispersion along AH and a significantly steeper gradient in AL direction (A) and a second band, which shows a steep dispersion and evolves rather similar in both high symmetry directions of the BZ (B).
Next we observe a shoulder suggesting the existence of a third band with a larger effective mass than B.
Our experimental results for $p$-polarized light are in qualitative agreement with previous ARPES studies at a photon energy of 11\,eV presented by Falke {\it et al.}\cite{Falke2021}
In the middle panel we repeat the same measurement with $s$-polarized light unveilling a third band (C), that is barely visible with $p$-polarized light.
Mapping the intensity difference between the two measurements, in the bottom panel we distinguish the bands with a significant out-of plane component or in plane nodal planes parallel to our analyzer slit ($k_{\rm y}$) (A and B) from the one with $k_{\rm x}$ in-plane components (C).\cite{Sugimoto2015,Moser2017}
\\
Comparing our experimental results with our DFT band structure calculations, we can identify the three bands predicted for the BZ sector we have investigated. 
Band positions as well as dispersion are remarkably well described.
In agreement with calculations reported by Strocov {\it et al.} \cite{Strocov2012} and Majchrzak {\it et al.} \cite{Majchrzak2021} we assign to the band A a V$_{\rm 3d}$ character.
The orbital character of band B is instead controversial. 
According to Strocov {\it et al.} it corresponds to the Se$_{\rm 4p_z}$ state, whereas the band structure calculations reported by Majchrzak {\it et al.} show two bands along AH, rather close to each other and strongly hybridizing, leading to a more $d$-like character close to the Fermi level and changing with increasing momentum.
The dispersion and symmetry of band C along $\Gamma$K agrees with the calculation reported by Strocov {\it et al.} for the Se$_{\rm 4p_{xy}}$ band.
In our data we see a clear dependence on the polarization of the incident light, which leads to the conclusion that bands B and C have different origin.
From the orbital projection of the band structure calculations ($cf.$ Fig.s \,S1 and S2 of the supplementary material), it becomes evident that band A is a pure V$_{\rm 3d}$ band, whereas band B and C have mixed character. 
Both bands keep partial $p$ flavor and can be distinguished by their original atomic orbital symmetry. 
The stronger-dispersing band B has $p_{\rm y}$ imprinting, whereas band C is more $p_{\rm x}$. 
This fits our experimental observation that B can be better probed with $p$-polarized light.

\begin{figure}[t]
\includegraphics[width=7.8cm]{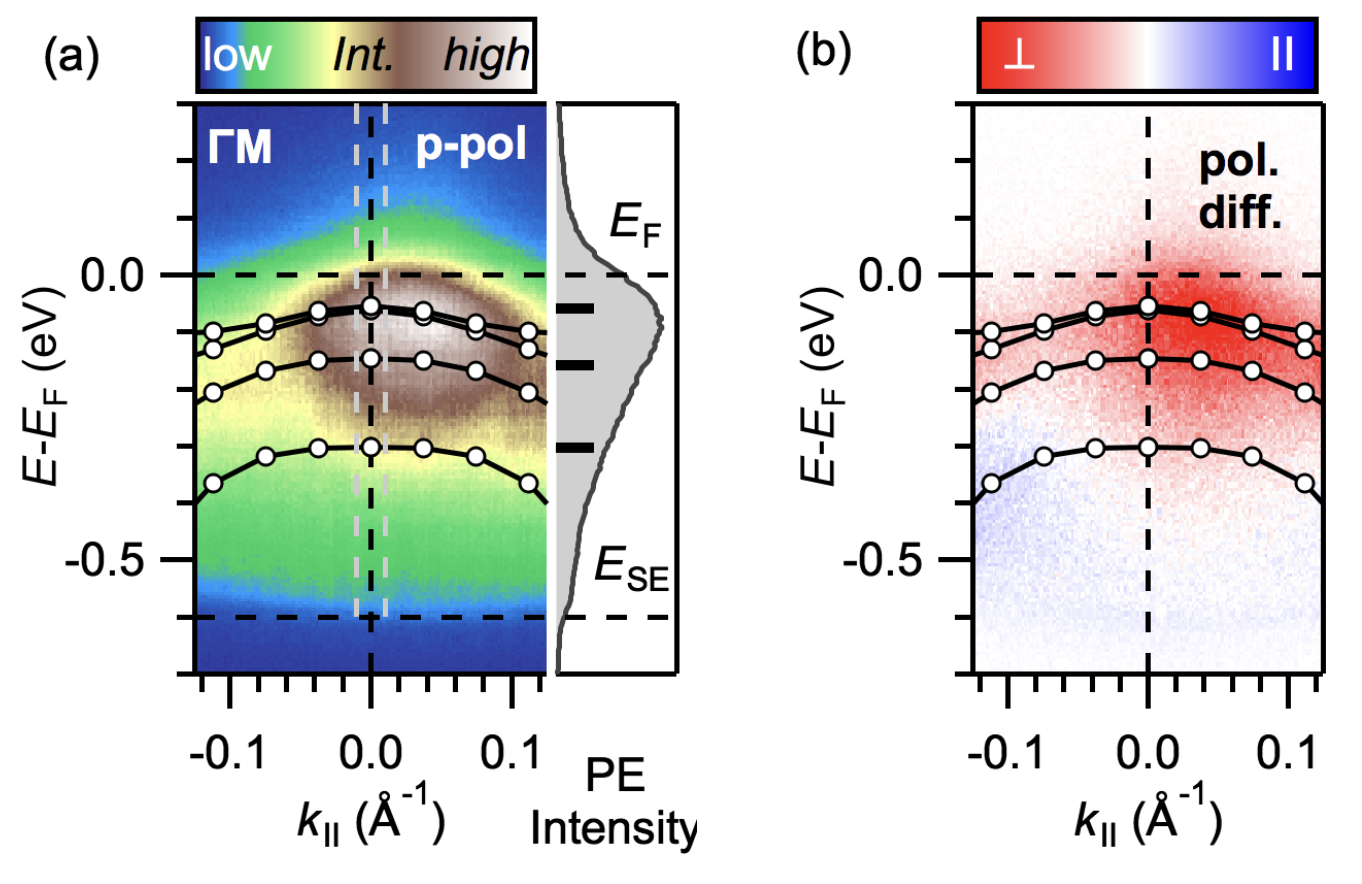}
\caption{Equilibrium ARPES measurements at 6.2\,eV in comparison with DFT band structure calculations along the $\Gamma$M direction. (a) 
Photoemission intensity distribution map as a function of binding energy and momentum parallel to the surface plane. The black curves with markers are showing the band structure calculations in the $\Gamma$M direction. At the right side of the map the energy distribution curve (EDC) integrated in a small window around zero momentum is shown. The black strips mark the calculated positions of the bands in the integration window of the EDC. (b) Polarization difference map ($p$-pol - $s$-pol) in comparison with the band positions derived from the calculations. $P$-polarized light gives more intensity for the hybridized Se and V bands close to the Fermi level.
}
\label{fig:6eV_data}
\end{figure}

Changing the probe photon energy to 6.2\,eV, we move from the AHL plane to the $\Gamma$KM plane of the BZ, if we consider the inner potential V$_0$ of 16\,eV as determined by Falke {\it et al.} \cite{Falke2021} from low photon energy synchrotron measurements in the range from 9 to 17\,eV. 
Instead, synchrotron measruements at photon energies above 65\,eV report inner potentials of only 7.5\,eV (\textit{cf.} Ref.\,[\onlinecite{Majchrzak2021}], supporting material) and Ref.\,[\onlinecite{Strocov2012}]).
The discrepancy between the two effective inner potential values for different incident photon energies could be based on final state effects. 
Smaller variations of V$_0$ can be observed over narrower photon energy ranges.\cite{Tamura1985} 
However, we assume that our photon energy of 6.2\,eV is sufficiently close to the photon energies used by Falke {\it et al.} and conclude that we probe the BZ very close to the $\Gamma$ point.
Note that this consideration does not include final state effects originating from resonant excitation at our probe energy.

The 6.2 eV photon energy reduces significantly our field of view in momentum space and binding energies. 
As we show in Fig.\,\ref{fig:6eV_data}, we observe photoemission intensity in a binding energy range of $E-E_{\rm F}\approx 0.6$\,eV.
From these data we can deduce the sample work-function $\Phi$ from the low energy cut-off ($E_{\rm SE}$) of the energy distribution curves (EDCs). 
We find $\Phi=h\nu-|E_{\rm SE}|=(5.64\pm 0.03)$\,eV, which is close to the value of $(5.76\pm0.05)$\,eV reported by Claessen {\it et al.}\cite{Claessen1990} and confirms a clean surface of good quality. 

The photoemission intensity increases towards the Fermi level. 
We observe a slight dispersion of the intensity maximum along $k_{\parallel}$. 
By comparing our band structure calculations with our experimental data, we find that three of the four valence bands are positioned in the binding energy range where we observe enhanced photoemission intensity. 
Instead, the fourth band is clearly separated from the region where we measure highest intensity. 
We mark the positions of the bands with black stripes in the EDC shown at the right side of Fig.\,\ref{fig:6eV_data}(a), which is taken at the $\Gamma$ point (\textit{cf.} integration window indicated by the dashed grey lines on top of the intensity distribution map).
With the given energy resolution we cannot resolve the bands crossing in this energy and momentum region.
For more clarity we report a polarization difference map in Fig.\,\ref{fig:6eV_data}(b), which is telling us that we observe higher intensity with $p$-polarized light (red) in the range of the upper three bands and slightly higher intensity with $s$-polarized light (blue) in the binding energy range of the fourth band.
The band with the highest binding energy has a Se$_{\rm 4p}$ character at the $\Gamma$-point, according to the calculations reported by Majchrzak {\it et al.}\cite{Majchrzak2021}, whereas the other three bands are strongly hybridizing, so that the two $p$-bands have also a significant $d$-character. 
This is in contrast to the calculations published by Jolie {\it et al.}\cite{Jolie2019}, but fits to our observation when comparing different probe polarizations.

\begin{figure*}[t]
\includegraphics[width=\textwidth]{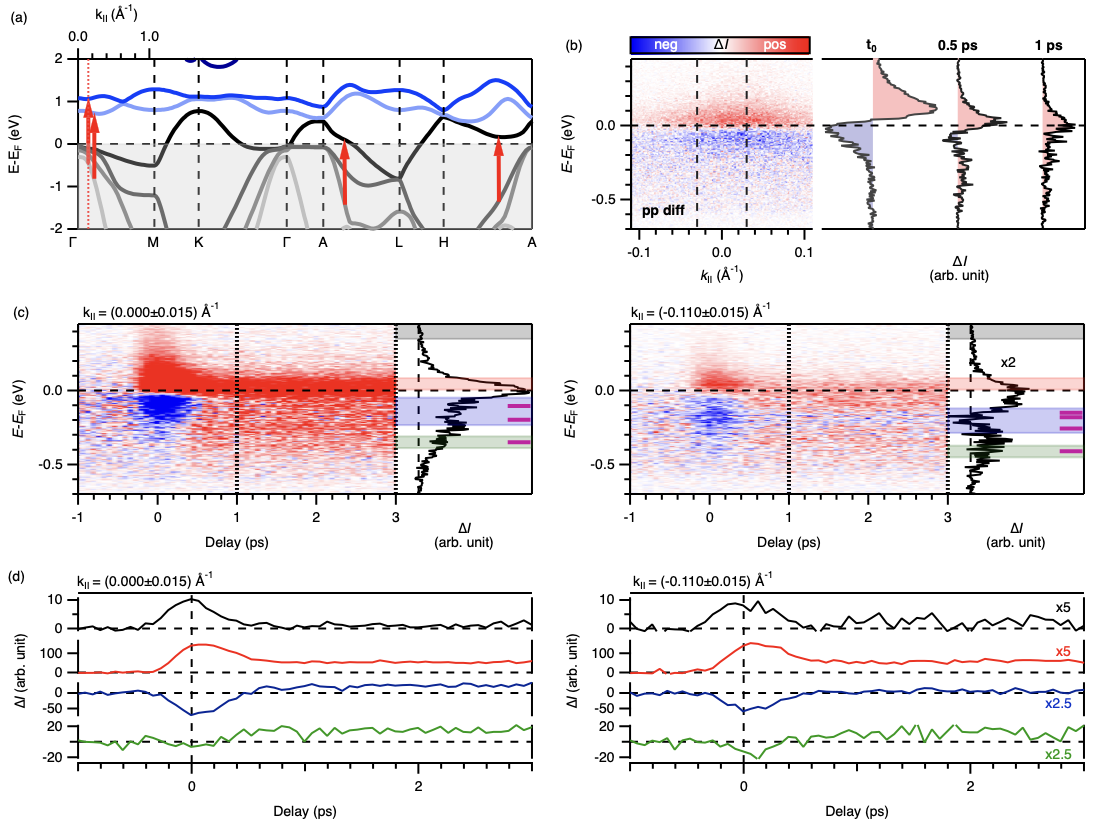}
\caption{Tr-ARPES measurements at 6.2\,eV probe energy. (a) Calculated band structure for 1$T$-VSe$_2$ bulk crystal structure in its normal phase. Red arrows mark $k$-values for resonant vertical transitions at our pump-energy of 1.55\,eV. (b) Pump-probe difference map for $p$-polarized probe pulses. At the right side of the panel three traces showing the spectral changes during the excitation (t$_0$) as well as 500\,fs and 1\,ps after the excitation are shown. The integration window is indicated in the map by the vertical lines. (c) Energy-resolved transient evolution of the photoemission intensity at two $k_{\parallel}$-values (0 and 0.11 \AA$^{-1}$) with an integration window of 0.03 \AA$^{-1}$. At the right side of the panels we show the difference spectra integrated between 1 and 3\,ps, showing the signature of the metastable state. (d) Selected (horizontal) time-cuts through the intensity maps recalling the color scheme of the colored areas in panel (c). For better visibility we multiplied the signal observed at higher $k$-values with the reported factors.}
\label{fig:6eV_data_tr}
\end{figure*}
\subsection{Pump-probe spectroscopy}
To investigate optically-induced changes in the electron density of states near the Fermi level, we performed pump-probe experiments using 1.55\,eV pump photon pulses with an incident fluence of $\approx 0.4$\,mJ/cm$^{2}$.
According to our band structure calculations shown in Fig.\,\ref{fig:6eV_data_tr}(a), there are selective $k_{\parallel}$ values along $\Gamma$M, AL and AH directions where resonant vertical transitions from bands with predominantly $p$-like character (grey) to $d$-like bands (black, blue) are predicted to be possible. 
The unoccupied bands around 1\,eV above the Fermi level are assigned to $d$-like character in agreement with Ref.\,[\onlinecite{Jolie2019}].\\
Out of equilibrium, we initially observe a decrease in the photoemission intensity below the Fermi level, indicating a depopulation of the bands (cf. the pump/non-pumped difference map shown in Fig.\,\ref{fig:6eV_data_tr}(b)) and an increase of the photoemission intensity just above the Fermi level.
This behaviour is expected for a broadening of the Fermi edge due to an increase of the electronic temperature by the pump pulse.
Indeed we are observing an almost symmetric shape of the difference EDC integrated in a small window around the $\Gamma$-point at the delay of temporal overlap between pump and probe pulses (t$_0$, cf. Fig.\,\ref{fig:6eV_data_tr}(b)).
At later delays, however, the difference EDC gets highly asymmetric, clearly showing a residual enhanced intensity above the Fermi level and even a net increase of the signal below the Fermi level in comparison to equilibrium conditions (cf. Fig.\,\ref{fig:6eV_data_tr}(b) for the EDC difference curves at $t=0.5$ and 1\,ps).\\
Tracing the evolution of the intensity distribution as a function of pump-probe delay at $\Gamma$ and at $k_{\parallel}=-0.11$\,\AA$^{-1}$ (see Fig.\,\ref{fig:6eV_data_tr}(c)), we observe an overall increase of the photoemission intensity in the range from -0.5\,eV up to the Fermi level in the whole momentum range we are investigating, surviving for several picoseconds (cf. Fig.\,\ref{fig:6eV_data_tr}(c)).
At the right side of the intensity difference maps in Fig.\,\ref{fig:6eV_data_tr}(c) we report the difference EDCs averaged over all pump-probe delays larger than 1\,ps, in order to display the spectral shape of the metastable state. 
At $\Gamma$ we observe an overall intensity increase, which reaches its maximum around the Fermi level and then seems to drop just below the Fermi level. 
We attribute this drop to a depopulation of the vanadium and selenium bands, whose central positions are marked as magenta stripes in Fig.\,\ref{fig:6eV_data_tr}(c).
This assignment is getting clearer when evaluating the difference EDCs at larger $k_{\parallel}$ values. 
The drop in intensity shifts here towards higher binding energies, according to the dispersion of the bands (cf. right panel of Fig.\,\ref{fig:6eV_data_tr}(c)).
\\
In Fig.\,\ref{fig:6eV_data_tr}(d) we show the temporal evolution of the photoemission intensity for four distinct energies, indicated by the colored stripes in panel (c).
The grey window is chosen in the uppermost part of our measurement window, well above the Fermi level, to extract a transient that should be influenced mainly by our pump-probe cross correlation and hence allows us to deduce t$_0$.
The red window describes the intensity evolution above the Fermi level, whereas the blue and the green windows are chosen to correspond to the positions of the valence bands.
In comparison to the grey curve, all other curves show a delayed reach of their maximum intensity change. 
A delayed depopulation of the valence bands is expected due to cascading effects, since the $k$-values for resonant excitation with 1.55\,eV are just out of our field of view.
At $k_{\parallel}=-0.11$\,\AA$^{-1}$ the intensity change is suppressed to close to zero in the range of the upper three valence bands.\\
Majchrzak {\it et al.}\cite{Majchrzak2021} were reporting a depletion of the Se$_{4p}$ band reaching a plateau within the first 500\,fs and not repopulating up to 5\,ps after photoexcitation. 
The authors were referring this to hot carriers affecting the low-energy soft phonon spectrum of the material and the electron-phonon coupling strength.
We conclude that the spectral shape of the difference EDCs results from an intensity increase in the whole range from -0.5\,eV up to the Fermi level, which is suppressed at some energies due to a long-lived depletion of the valence band population after photoexcitation.
In contrast to Majchrzak {\it et al.},\cite{Majchrzak2021} we do not observe a net reduction of the photoemission intensity with respect to the equilibrium intensity due to a strong depopulation of the V$_{3\rm d}$ band at the $\Gamma$ point that persists for several ps. 
We attribute this to the lower excitation fluence we are using in our experiments, which allows us to observe the net increase of the photoemission intensity.
We will discuss possible origins of this phenomenon at the end of this section.

\begin{figure*}[t]
\includegraphics[width=\textwidth]{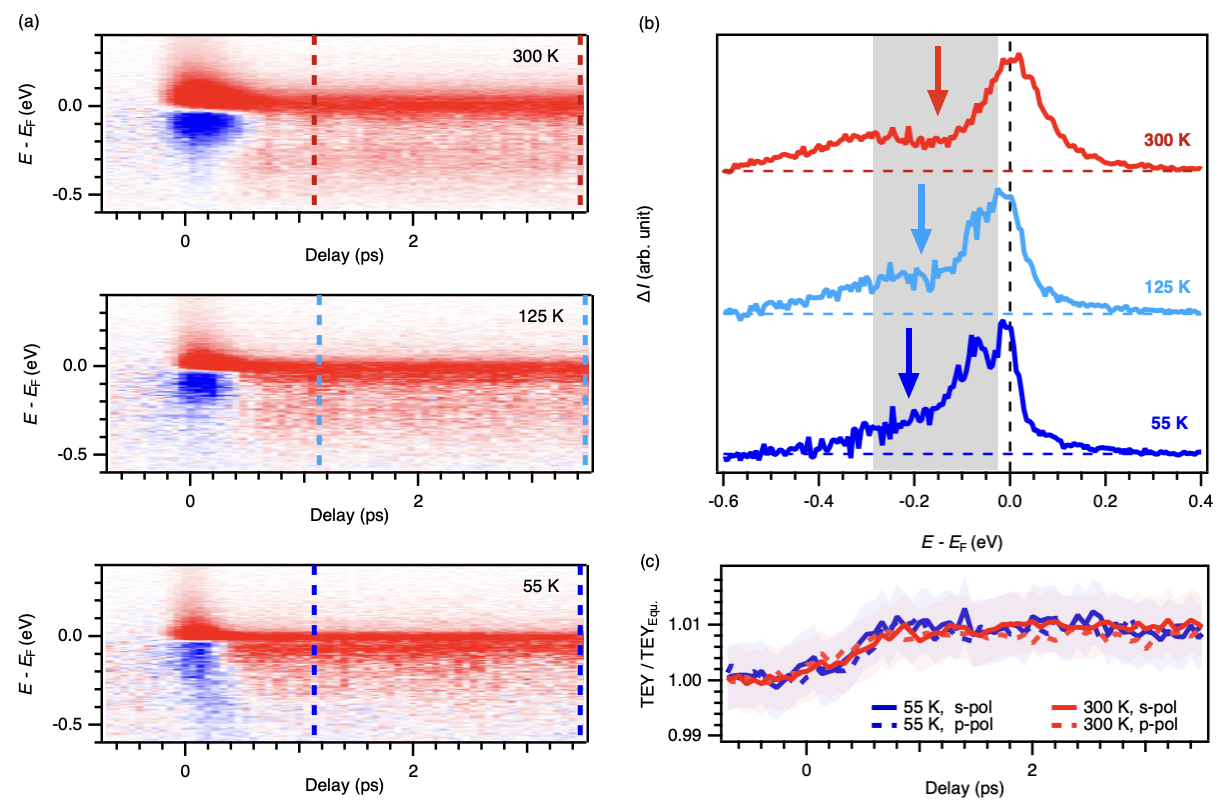}
\caption{(a) Angle-integrated tr-ARPES data recorded with $s$-polarized probe light at 300\,K, 125\,K and 55\,K. 
(b) Difference EDCs of the metastable density of states modification extracted from the difference intensity maps shown in panel (a) between the indicated dashed lines. 
The grey shaded area indicates the region where we observe a suppression of the signal due to the depopulation of the crossing valence bands at room temperature. 
We use the arrow to indicate that the region where the signal is getting suppressed shifts to higher binding energies when decreasing the temperature and switching to the charge density wave phase.
(c) Change in the total electron yield (TEY) as a function of pump-probe delay. 
The yield is integrated over our full measurement window (momentum and energy space). 
Independently of the sample temperature and probe light polarization, the yield increases by 1\,\% upon pumping.}
\label{fig:6eV_data_tr_Temp}
\end{figure*}
In Fig.\,\ref{fig:6eV_data_tr_Temp} we compare the results at room temperature (well above T$_{\rm c}$) discussed so far, with measurements performed at sample temperatures of 125\,K (slightly above T$_{\rm c}$) and 55\,K (well below T$_{\rm c}$).
Here we integrated over the whole momentum range from -0.1 to 0.1\,\AA$^{-1}$.
Besides the sharpness of the Fermi edge we do not observe any significant difference between the measurements acquired at sample temperatures of 300\,K and 125\,K. 
Cooling well below T$_{\rm c}$ (55\,K) we find a more pronounced delayed band depletion at binding energies of the order of 0.5\,eV.
After 1\,ps, also in this case we find the same broad-range intensity increase as above T$_{\rm c}$. 
Comparing the difference EDCs for the three different temperatures (cf. Fig.\,\ref{fig:6eV_data_tr_Temp}(b)), we observe a clear downshift of the onset of the signal suppression due to valence band depletion with decreasing sample temperature.
We interpret this observation, as well as the retarded depletion at the higher binding energies, as indications of slight changes in the valence bands dispersion when lowering the sample temperature.
We consider these changes in the band structure as related to the CDW phase and conclude that with the given pump fluence the CDW is at least partially preserved.
The optically-induced intensity increase however does not seem to depend on the crystals phase.
As shown in Fig.\,\ref{fig:6eV_data_tr_Temp}(c), we evaluate the total electron yield (TEY) integrated in the complete field of view we are probing with 6.2\,eV photons. 
We observe an increase by 1\,\% of the equilibrium TEY upon 1.55\,eV pumping. 
This observation is identical when starting from the normal or the CDW phase and does not depend on the probe polarization. 
\\\\
Summing up our observations, we exclude that the observed intensity increase is related to a closure of the CDW gap due to optical excitation. 
The signal shows no temperature dependence and our measurements at 300\,K are far above the critical temperature of 110\,K.
We also exclude the signal originating from a shift of the entire band structure. 
1.55\,eV pump and HHG probe measurements published by Majchrzak {\it et al.} show no indication of such a shift at higher excitation fluences.\cite{Majchrzak2021} Since optically-induced band shifts should scale with the fluence, larger shifts are not expected at lower fluence.
Furthermore, we demonstrated that our positive signal decreases following the equilibrium dispersion of the selenium bands (cf. Fig.\,\ref{fig:6eV_data_tr}(c)). 
At larger $k_{\parallel}$ values the data of Majchrzak {\it et al.} show an indication of a drastic shift of the V$_{\rm 3d}$ band only, whose origin however is not discussed.
Since our static data (cf. Fig.\,\ref{11eV_ARPES} and Fig.\,\ref{fig:6eV_data}) clearly show that the V$_{\rm 3d}$ is probed with $p$-polarized light only, we also exclude a shift in the vanadium band position to be the origin of our signal, since we can observe it likewise with both probe polarizations.
\\
A recent article claimed that a flat band is present in large parts of the bulk VSe$_2$ BZ and can be shifted downwards via electron doping through donor adatoms adsorbed on the sample surface.\cite{Yilmaz2022}
These authors speculated that polaron formation could be at the origin of this mechanism.
Pump-induced electron doping could be an explanation for our observation. 
However, so far theoretical confirmation of flat bands or polarons in bulk VSe$_2$ is lacking.
\\
Picosecond-scale modifications of the electronic density of states after optical excitation could also result from trapping electrons at defects in the crystal structure, as it was discussed {\it e.g.} for MoS$_2$ monolayers.\cite{Wang2015}
In our case of increased intensity below the Fermi level, this would mean that the defect levels are dragged below the Fermi level upon population.
\\
An additional explanation of the metastable signal can be related to the light-induced suppression of electronic correlation in 1{\it T}-VSe$_2$, resulting in the recovery of additional density of states in the energy region of the vanadium 3d bands. Indeed, Biswas {\it et al.}\cite{Biswas2021} reported that significant electronic correlation is present in VSe$_2$. In particular, the fingerprint of this effect was found in the large broadening of the vanadium 3d states below the Fermi level, which could only be captured by dynamical mean field theory calculations with a large Hubbard interaction strength of the order $U = 6$\, eV. Similarly to what reported in \cite{Cilento2018}, when energy is provided to the system via photoexcitation, the strength of electronic correlations can be markedly reduced, hence new electronic states in the energy region of the vanadium 3d bands, that were suppressed by localization, can emerge, similarly to the collapse of a correlation gap in photoexcited Mott insulators.
\\
Whichever is the origin of the photoinduced metastable electronic state with enhanced density of states that we revealed, the large increase of the carrier density at the Fermi level implies a transient enhanced metallicity of the material.

\section{summary and conclusion}
In this study we used polarization-dependent ARPES to disentangle the contributions of different vanadium- and selenium-derived bands to the valence band structure close to the $\Gamma$ and A points of the 1{\it T}-VSe$_2$ hexagonal BZ.
Our results show good agreement with theoretical band structure calculations, allowing for one-to-one assignments of the probed bands. 
Furthermore, we used our extracted knowledge about the position and dispersion of different bands to interpret the changes in the electronic density of states out-of-equilibrium. 
We detect the formation of a metastable configuration of the density of states reached after optical pumping, leading to additional electron density of states spread over 0.5\,eV below the Fermi level, which has not been observed so far.
This finding asks for further investigation. 
Nonetheless, the possibility to alter the electronic bands by ultrashort light pulses in a way not attainable under equilibrium paves the way to the design of new metastable states of matter with tailored electronic properties and novel functionalities, connected to the light-induced enhanced metallicity of the system, under the form of an enhanced density of states extending from the valence-band region up to the Fermi level.

\begin{acknowledgments}
W.B. acknowledges financial support from the German Academic Exchange Service (DAAD) for financing a research stay at the Institute of Atomic and Molecular Sciences at Academia Sinica in Taipei (Taiwan) for establishing collaborations.
\end{acknowledgments}

\bibliographystyle{apsrev4-1}
%\bibliography{VSe2}


\end{document}

\title[Supplementary Material: Ultrafast Band Structure Dynamics in Bulk 1{\it T}-VSe$_2$]
{Supplementary Material: Ultrafast Band Structure Dynamics in Bulk 1{\it T}-VSe$_2$}

\author{Wibke Bronsch}
\email[Corresponding author: ]{wibke.bronsch@elettra.eu}
\affiliation{Elettra - Sincrotrone Trieste S.C.p.A., Strada Statale 14 - km 163.5 in AREA Science Park, 34149 Basovizza, Trieste, Italy}

\author{Manuel Tuniz}
\affiliation{Dipartimento di Fisica, Universit\`{a} degli Studi di Trieste, 34127 Trieste, Italy}

\author{Denny Puntel}
\affiliation{Dipartimento di Fisica, Universit\`{a} degli Studi di Trieste, 34127 Trieste, Italy}

\author{Alessandro Giammarino}
\affiliation{Dipartimento di Fisica, Universit\`{a} degli Studi di Trieste, 34127 Trieste, Italy}

\author{Fulvio Parmigiani}
\affiliation{Dipartimento di Fisica, Universit\`{a} degli Studi di Trieste, 34127 Trieste, Italy}
\affiliation{Elettra - Sincrotrone Trieste S.C.p.A., Strada Statale 14 - km 163.5 in AREA Science Park, 34149 Basovizza, Trieste, Italy}
\affiliation{International Faculty, University of Cologne, Albertus-Magnus-Platz, 50923 Cologne, Germany}

\author{Yang-hao Chan}
\affiliation{Institute of Atomic and Molecular Sciences, Academia Sinica, Taipei 10617, Taiwan}
\affiliation{Physics Division, National Center of Theoretical Physics, Taipei 10617, Taiwan}

\author{Federico Cilento}
\email[Corresponding author: ]{federico.cilento@elettra.eu}
\affiliation{Elettra - Sincrotrone Trieste S.C.p.A., Strada Statale 14 - km 163.5 in AREA Science Park, 34149 Basovizza, Trieste, Italy}

\date{\today}

\maketitle

\section{Projected band structure calculation}
\begin{figure}[h!]
\includegraphics[width=\textwidth]{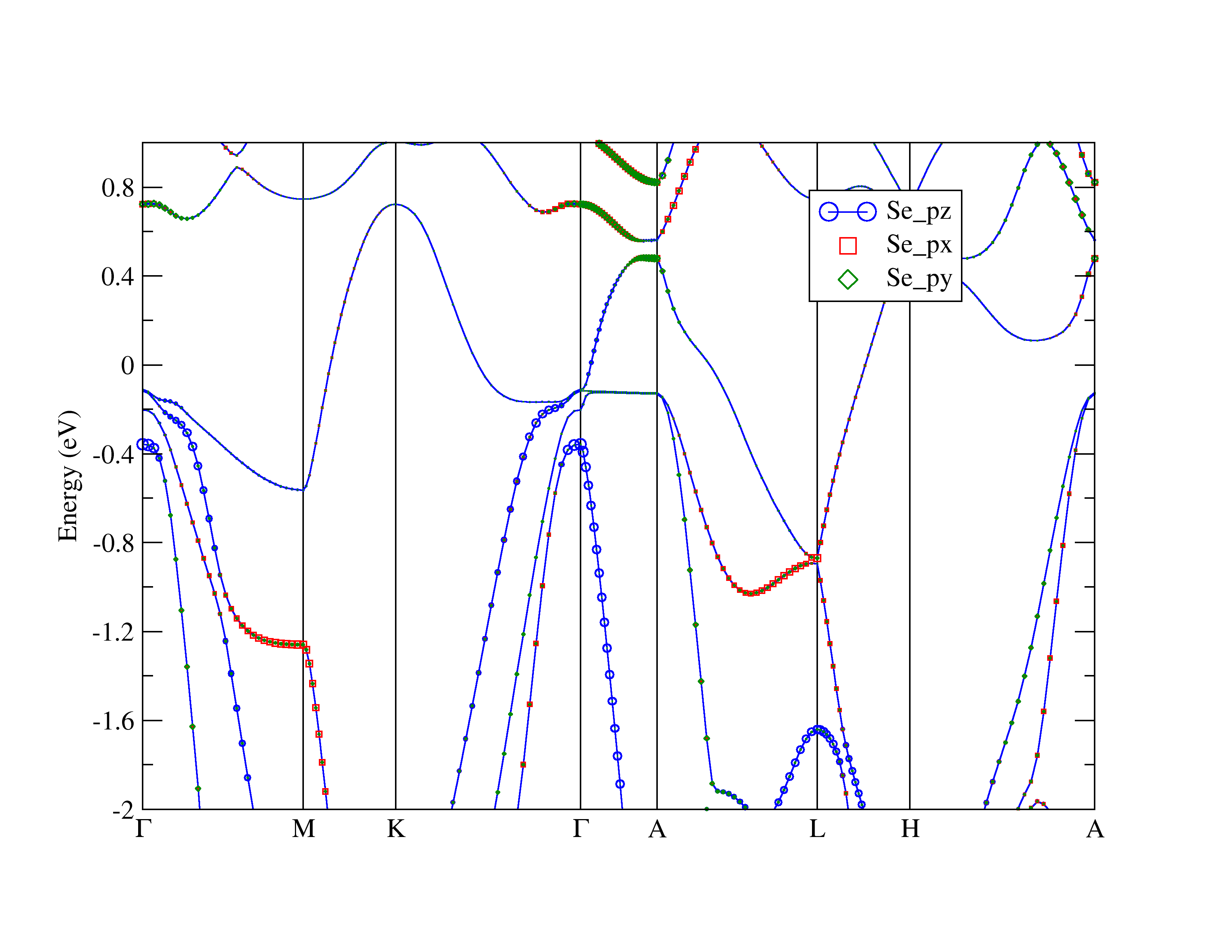}
\caption{
Projected band structure calculations considering the Se 4p orbital characters.
The size of the symbols in the plot corresponds to the relative amount of the corresponding orbital character.}
\label{fig:Se4p_projection}
\end{figure}

\begin{figure}[t]
\includegraphics[width=\textwidth]{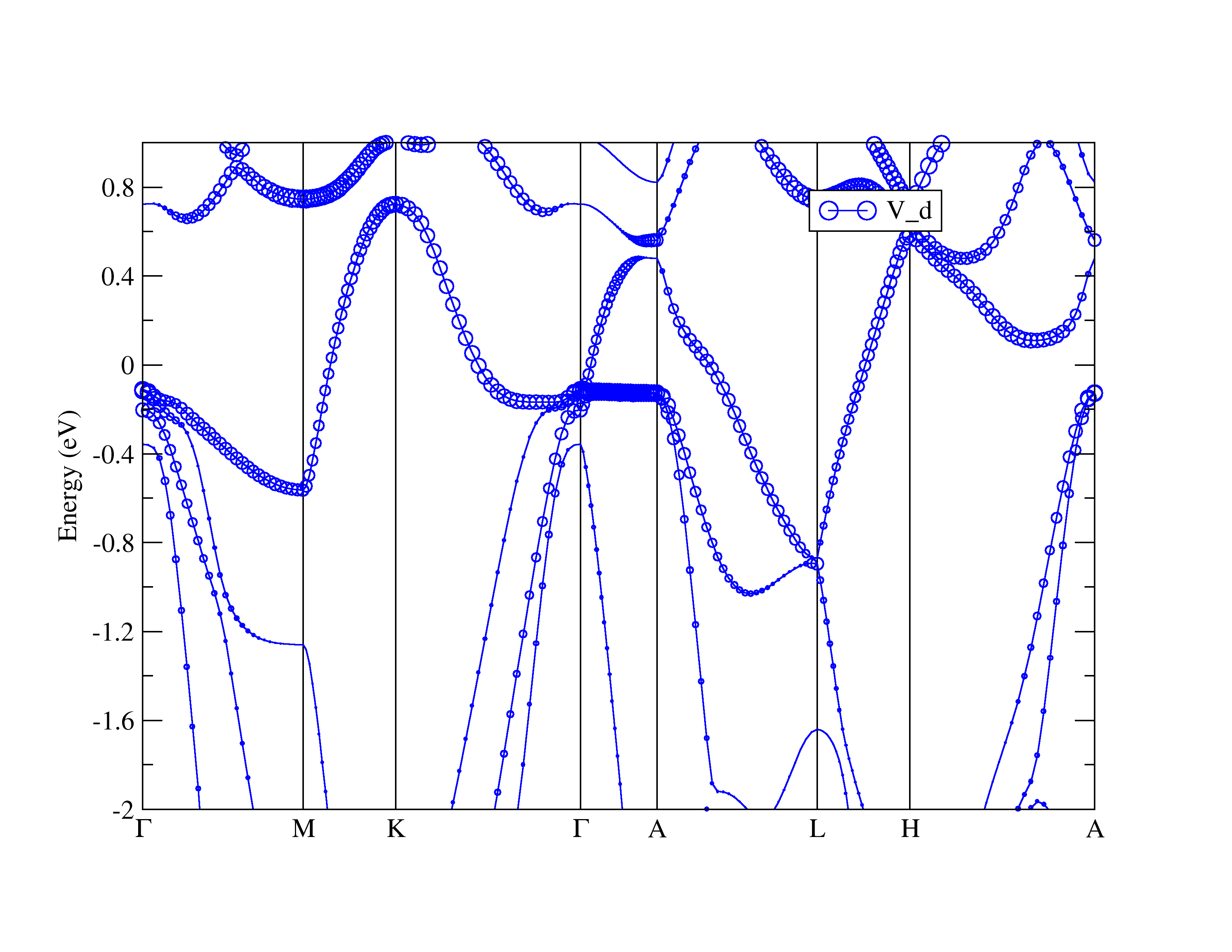}
\caption{
Projected band structure calculations considering the V 3d orbital characters.
}
\label{fig:V3d_projection}
\end{figure}

Additionally to the band structure calculations shown in the main text, we report here separately the band structure projections for $p$ (\textit{cf.} Fig.\,\ref{fig:Se4p_projection}) and $d$ \textit{cf.} Fig.\,\ref{fig:V3d_projection}) orbitals.
In vicinity of the Fermi level all bands are getting significant d character due to hybridization effects. 
The $p_{\rm z}$ band which can be found at $\Gamma$ at a binding energy of about 350\,meV shows strong dispersion along $k_{\rm z}$ (compare $\Gamma A$ direction in Fig.\,\ref{fig:Se4p_projection}) and is located clearly outside of our field of view for the measurements in the A plan (photoemission data acquired with a probe photon energy of 10.8\,eV).
The two remaining bands have partial $p_{\rm x}$ and $p_{\rm y}$ character, which makes them distinguishable with $s$- and $p$-polarized light in our measurement geometry.\cite{Sugimoto2015,Moser2017}

%\clearpage
\section{Joined density of states}
\begin{figure}[t]
\includegraphics[width=\textwidth]{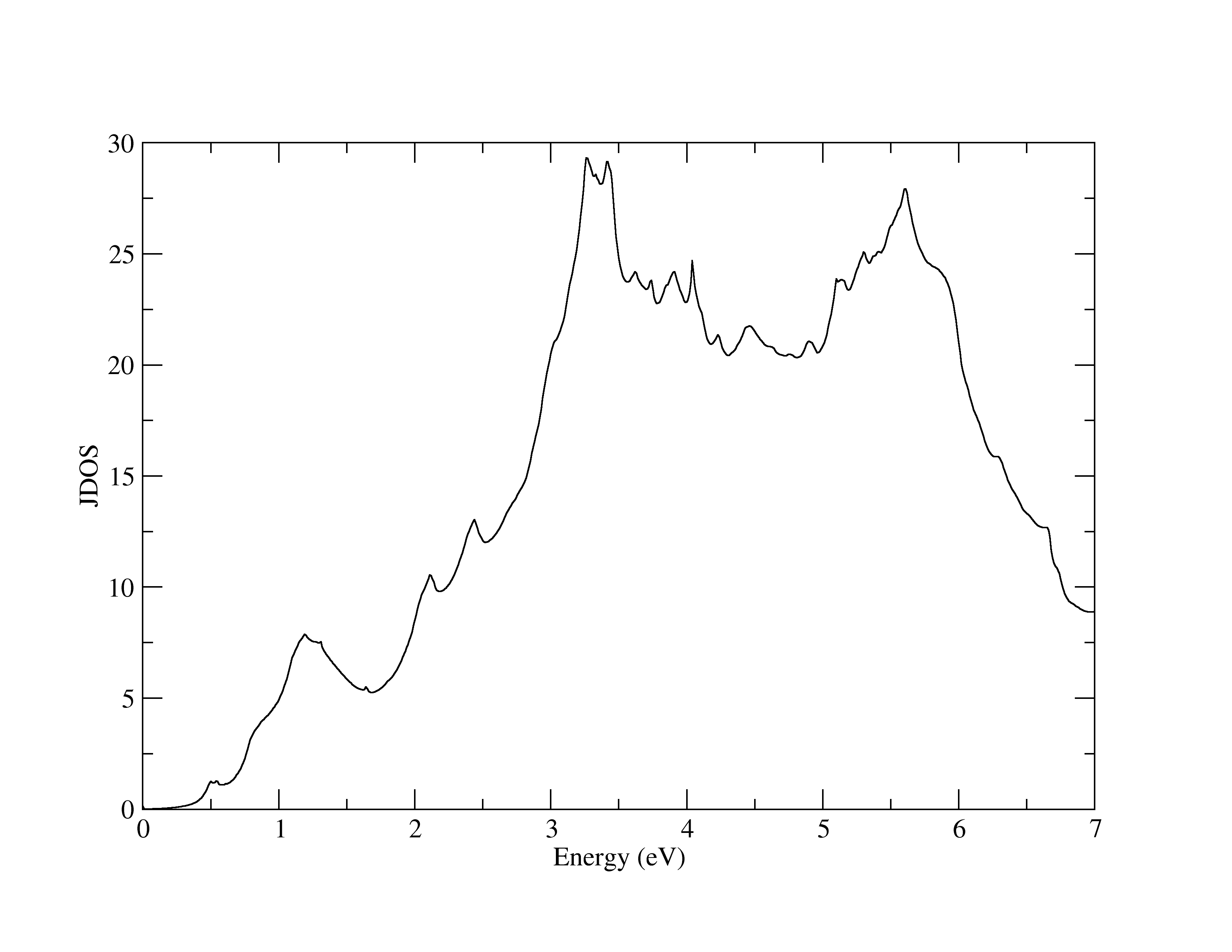}
\caption{
Joint Density of States (JDOS) for VSe$_2$ in units of states/eV, as computed from the band structure calculations reported in the main text.
}
\label{fig:jdos}
\end{figure}

Fig. \ref{fig:jdos} shows the Joint Density Of States (JDOS) for optical transitions of VSe$_2$. It is calculated over a large photon energy interval h$\nu$=0-7 eV, using as an input the band structure calculations reported in the main text. The JDOS indicates the number of pairs of electronic states connected by an optical transition for photons at energy h$\nu$, that excite electrons from occupied to unoccupied states. Hence, it provides the information of states available for optical transitions.

In connection to our experiments, we are interested to the value of the JDOS at the pump photon energy h$\nu_{pump}$=1.55 eV.

